\documentclass{PoS}

\usepackage{amsmath,amsfonts,amssymb,booktabs}
\title{Isospin Effects by Mass Reweighting}

\ShortTitle{Isospin Effects by Mass Reweighting}
        
\author{\speaker{Jacob Finkenrath}$\;^{,a}$, Francesco Knechtli $^{a}$, B\"orn Leder $^{a,b}$
	\\
      $^{a}$ Department of Physics, Bergische Universit\"at Wuppertal\\
      Gaussstr. 20, D-42119 Wuppertal, Germany\\
      $^{b}$ Department of Mathematics, Bergische Universit\"at Wuppertal\\
      Gaussstr. 20, D-42119 Wuppertal, Germany\\
      E-mail: \email{finkenrath@physik.uni-wuppertal.de}}

\abstract{Most of today's lattice simulations are performed in the isospin symmetric
limit of the light quark sector. Mass reweighting is a technique to
include effects of isospin breaking in the sea quarks at moderate numerical cost.
We will give a summary of our recent results on fine lattices with light quark masses
and will show how light quark masses can be extracted 
by introducing suitable tuning conditions for the bare mass parameters.

In general the reweighting factor introduces additional fluctuations and thus increases the statistical
uncertainties. In the case of isospin reweighting this factor is a ratio of fermion
determinants. The stochastic evaluation of the determinants potentially leads to stochastic noise 
in observables. We show the quark mass and the volume dependence of these fluctuations.
}

\FullConference{The 32nd International Symposium on Lattice Field Theory\\
                 23-28 June, 2014\\
                 Columbia University New York, NY}

\begin{document}

\section{Introduction}

Mass Reweighting \cite{Hasenfratz:2008fg} is an interesting and efficient method to correct and to include
effects of quark masses. It can be used for tuning, e.g.~the strange quark mass $m_s$ in a 2+1 simulation or
the isospin splitting in the up-quark mass $m_u$ and the down-quark mass $m_d$. Moreover it can be applied to understand the mass behavior
of observables. For small corrections it is applicable and more efficient than new simulations.
Mass reweighting involves the evaluation of fermion determinants
which can be rewritten by an integral representation. This integral representation
can be estimated by Monte Carlo integration which needs around 100 inversions of the Dirac operator
to control the stochastic noise efficiently.

The reweighting factor enters the measurement of an observable by \cite{Ferrenberg:1988yz}
\begin{equation}
    \langle O \rangle_W = \frac{\langle O  W \rangle}{\langle W \rangle} = \langle O  \tilde{W} \rangle 
    \label{eq:Obre}
\end{equation}
where the mass reweighting factor for $n_f$ flavors of quarks
\begin{equation}
W= \prod_{i=1}^{n_f} \frac{\det D_{m_{new,i}}}{\det D_{m_{old,i}}}
\end{equation}
is normalized with $ \tilde{W} =  W / \langle W \rangle$.
Here, the Dirac operator is given by the clover improved Wilson Dirac operator $D_m = D + m$.
The reweighting factor can be rewritten as a determinant of a ratio matrix $M$ with
\begin{equation}
W= \frac{1}{\det \prod_{i=1}^{n_f} \left[D^{-1}_{m_{new,i}} D_{m_{old,i}}\right]} = \frac{1}{\det M}.
\end{equation}
In general lattice simulations are done in the isospin symmetric limit in the light quark sector by setting the
light quark masses to the average light quark mass $m_{ud}= 0.5 (m_u + m_d)$.
The idea is to use mass reweighting to introduce isospin breaking.
The reweighting is performed by splitting up the light quark masses by keeping the average quark mass constant $2 m_{ud} = m_u+m_d =const$
and it follows with the mass shift $\Delta m_{ud} = m_d -m_u$
\begin{equation}
         m_{u} = m_{ud} - 0.5 \cdot \Delta m_{ud} \leftarrow \quad m_{ud} \quad  \rightarrow  m_{ud} + 0.5 \cdot \Delta m_{ud} =  m_d.
\end{equation}
This leads to the isospin reweighting factor
\begin{equation}
W_{iso}= \frac{1}{\det \left[D^{-1}_{m_u} D_{m_{ud}} D^{-1}_{m_d} D_{m_{ud}}\right]} = \frac{1}{\det M_{iso}}.
\label{eq:isore}
\end{equation}
Now, the determinant of the non--hermitian matrix $M$ can be rewritten by an integral representation
given by
\begin{equation}
\frac{1}{\det M} = \int \textrm{D}[\eta] \exp\{-\eta^\dagger M \eta\}
\label{eq:ingr}
\end{equation}
which holds for $\lambda (M+M^\dagger)>0$ \cite{Finkenrath:2013soa} and 
the normalized integral measure is given by ${\textrm{D}}[\eta] = \prod_{j=1}^n \textrm{d}x_j \textrm{d}y_j / \pi$ 
with $\eta_j= x_j+ i y_j$. The integral eq.~\eqref{eq:ingr} can be estimated stochastically
\begin{equation}
\frac{1}{\det M} = \frac{1}{N_{\eta}} \sum_{i=1}^{N_{\eta}} e^{-\eta_i^\dagger(M - I)\eta_i} + \mathcal{O}(1/\sqrt{N_\eta}) 
\end{equation}
by drawing $N_\eta$ pseudofermion fields $\eta$ distributed via the normalized function $\propto \exp\{- \eta^\dagger \eta\}$
and $I$ the unit matrix with dimension of $M$.
Note for every drawn field $\eta$ $n_f$--inversions of the Wilson Dirac operator $D_{m_i}$ have to be performed.

In general mass reweighting introduces fluctuations which increase the statistical error.
These fluctuations are the ensemble fluctuations, introduced by the ensemble average in eq.~\eqref{eq:Obre},
and the stochastic fluctuations, introduced by the stochastic estimation of the integral eq.~\eqref{eq:ingr}.
The fluctuations are given by the variance averaged over the ensemble and the pseudofermions $\eta$.
We will define the variance of the integral representation eq.~\eqref{eq:ingr} by 
$\sigma^2 = \langle \langle w w^\dagger \rangle_\eta \rangle  - \langle \langle w \rangle_\eta \rangle \langle \langle w^\dagger \rangle_\eta \rangle$
with the stochastic estimate $w(U,\eta) = 1/N_\eta \sum_i \exp\{- \eta_i^\dagger (M-I) \eta_i \}$.
By performing the $\eta$--average $\langle \rangle_\eta$ (i.e.~all $\eta_i$ independently) the fluctuations are given for finite $N_\eta$ by
\begin{equation}
 \sigma^2 = \frac{1}{N_\eta} \left\langle \frac{1}{\det ( M + M^\dagger - I) } \right\rangle + \frac{N_\eta-1}{N_\eta}  \left\langle \frac{1}{\det M M^\dagger } \right\rangle  - \left\langle \frac{1}{\det M} \right\rangle \left\langle \frac{1}{\det M^\dagger } \right\rangle
 \label{eq:fl}
\end{equation}
which holds for $\lambda (M+M^\dagger-I)>0$.
The stochastic fluctuations for one configuration are given by
neglecting the ensemble average $\langle \rangle$ and vanish for $N_\eta \to \infty$. 
Moreover by introducing a mass interpolation between the start and the target mass
the stochastic fluctuations can be further controlled, i.e.~if no Dirac operator
has a zero eigenvalue during the interpolation the condition $\lambda (M+M^\dagger)>1$
can be insured \cite{Leder:2014ota}. In this case the number of inversions is $N_{inv} \propto N\cdot N_\eta$,
where $N$ is the number of interpolation steps. Note for many reweighting cases
it is effecient to use the even-odd preconditioned Wilson--Dirac operator (e.g.~see \cite{Finkenrath:2012cz}),
however we do not find an improvement in the case of isospin reweighting.
The ensemble fluctuations can be tamed by including additional quarks
into the reweighting process, e.g.~in the case of isospin reweighting the fluctuations
are minimized by keeping the average quark mass $m_{ud}=0.5(m_u+m_d)$
constant during the reweighting.

\begin{table}
\begin{center}
\caption{The table shows the analyzed CLS - ensembles generate with 
$n_f = 2$ - $\mathcal{O}(a)$ improved Wilson fermions with $m_{ud} = m_u = m_d$ \cite{Fritzsch:2012wq}.
We used ensembles of three different lattice spacings $a$ with pion masses $m_\pi$ from $580$ MeV down to $192$ MeV and lattice volume $V/a^4$. 
The number of configurations $N_{cnfg}$ are seperated by MDUs/config with $R_{act}$ the relative number of active links.
The maximal reweighting range is given by the quark mass shift in the $\overline{MS}$-scheme $\Delta m_{R,max}$. 
The renormalized mass is defined as in \cite{Fritzsch:2012wq}.}
\begin{tabular}{ccccrcc}
\toprule
ID & $V/a^4$          & $a$ [fm] & $m_\pi$ [MeV] & $N_{cnfg}$ & MDUs/config $\cdot R_{act}$  & $\Delta m_{R,max}$ [MeV] \\
\midrule
A5 & $\phantom{0}64\times 32^3$  & 0.076    & 330           & 202        &  $\phantom{0}20\cdot1\phantom{.37}$              &  $4.43(60)$ \\
\midrule
E4 & $\phantom{0}64\times 32^3$  & 0.066    & 580           & 100        &  $\phantom{0}16\cdot0.37$     &  $7.1(17)$  \\
D5 & $\phantom{0}48\times 24^3$  &   ``     & 440           & 503        &  $\phantom{00}8\cdot1\phantom{.37}$ &  $5.9(10)$  \\
E5 & $\phantom{0}64\times 32^3$  &   ``     & 440           & 99         &  $160\cdot0.37$    &  $6.01(96)$ \\
F7 & $\phantom{0}96\times 64^3$  &   ``     & 270           & 350        &  $\phantom{0}16\cdot0.37$     &  $5.01(38)$ \\
G8 & $128\times 64^3$            &   ``     & 192           & 90         &  $\phantom{00}8\cdot1\phantom{.37}$               &  $5.86(56)$ \\
\midrule
O7 & $\phantom{0}96\times 48^3$  & 0.049    & 270           & 98         &  $\phantom{0}40\cdot1\phantom{.37}$              &  $5.63(40)$ \\
\bottomrule
\end{tabular}
\label{tab:ens}
\end{center}
\end{table}

Here, we will discuss mass reweighting by introducing an isospin breaking in the light quarks.
We will show the scaling of the different fluctuations (ensemble and stochastic)
and how the up-- and down--quark mass can be extracted from the analyzed ensembles (see tab.~\ref{tab:ens}).

\section{Isospin Reweighting}

\begin{figure}
\begin{center}
\includegraphics*[angle=0,width=0.46\textwidth]{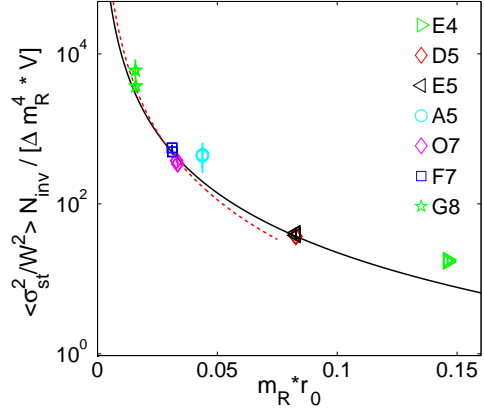}
\caption{\small The relative stochastic fluctuations of the isospin reweighting factor is shown as
a function of the renormalized quark mass. $\sigma^2_{st}$ is the average of the stochastic variance estimated using $N_\eta = 6$.}
\label{fig:sflc}
\end{center}
\end{figure}

The isospin reweighting factor eq.~\eqref{eq:isore} can be expanded in $\Delta m_{ud}^2$
\begin{equation}
  W_{iso} = \frac{\det D_{m_u}\det D_{m_d}}{\det D_{m_{ud}}^2} = 1 + \Delta m_{ud}^2 \textrm{Tr} (D_{m_{ud}}^{-2}) + \mathcal{O}(\Delta m_{ud}^4)
\label{eq:Wiso}
\end{equation}
by using $\det M = \exp \{ \textrm{Tr}(\ln(M))\}$. 
The fluctuations $\sigma^2$ in eq.~\eqref{eq:fl} of the isospin reweighting factor can be expanded in $\Delta m_{ud}^2$.
It can be shown that the stochastic fluctuations $\sigma_{st}^2$ decouple from the ensemble fluctuations $\sigma^2_{ens}$
with $\sigma^2 = \sigma_{st}^2 + \sigma_{ens}^2$.
The stochastic fluctuations in the isospin reweighting case are given by
\begin{equation}
  \left\langle\frac{\sigma^2_{st}(N_{inv})}{W^2}\right\rangle = \frac{\Delta m_{ud}^4}{N_{inv}} \; \left\langle \textrm{Tr} \frac{1}{\left(D_{m_{ud}}D_{m_{ud}}^\dagger\right)^2} \right\rangle + \mathcal{O}\left(\Delta m_{ud}^6\right).
\end{equation}
The ensemble fluctuations are
\begin{equation}
\frac{\sigma^2_{ens}}{\langle W \rangle^2} =  \Delta  m_{ud}^4 \; \textrm{var}\left(\textrm{Tr} \left[ \frac{1}{D_{m_{ud}}^{2}} \right]\right) + \mathcal{O}\left(\Delta m_{ud}^6\right)
\label{eq:sens}
\end{equation}
with the variance $\textrm{var}(O)=\left\langle O^2 \right\rangle - \langle O \rangle^2$.
Note this is true because the Dirac operator is $\gamma_5$-hermitian.
Now, the cost can be derived by demanding that the stochastic fluctuations do not dominate the
ensemble fluctuations, e.g.~that $ \sigma^2_{st}(N_{inv})/\sigma^2_{ens} \overset{!}{\sim} 0.1$.

\begin{figure}[t]
\begin{minipage}[c]{.46\textwidth}
  \begin{center}
   \includegraphics*[angle=0,width=\textwidth]{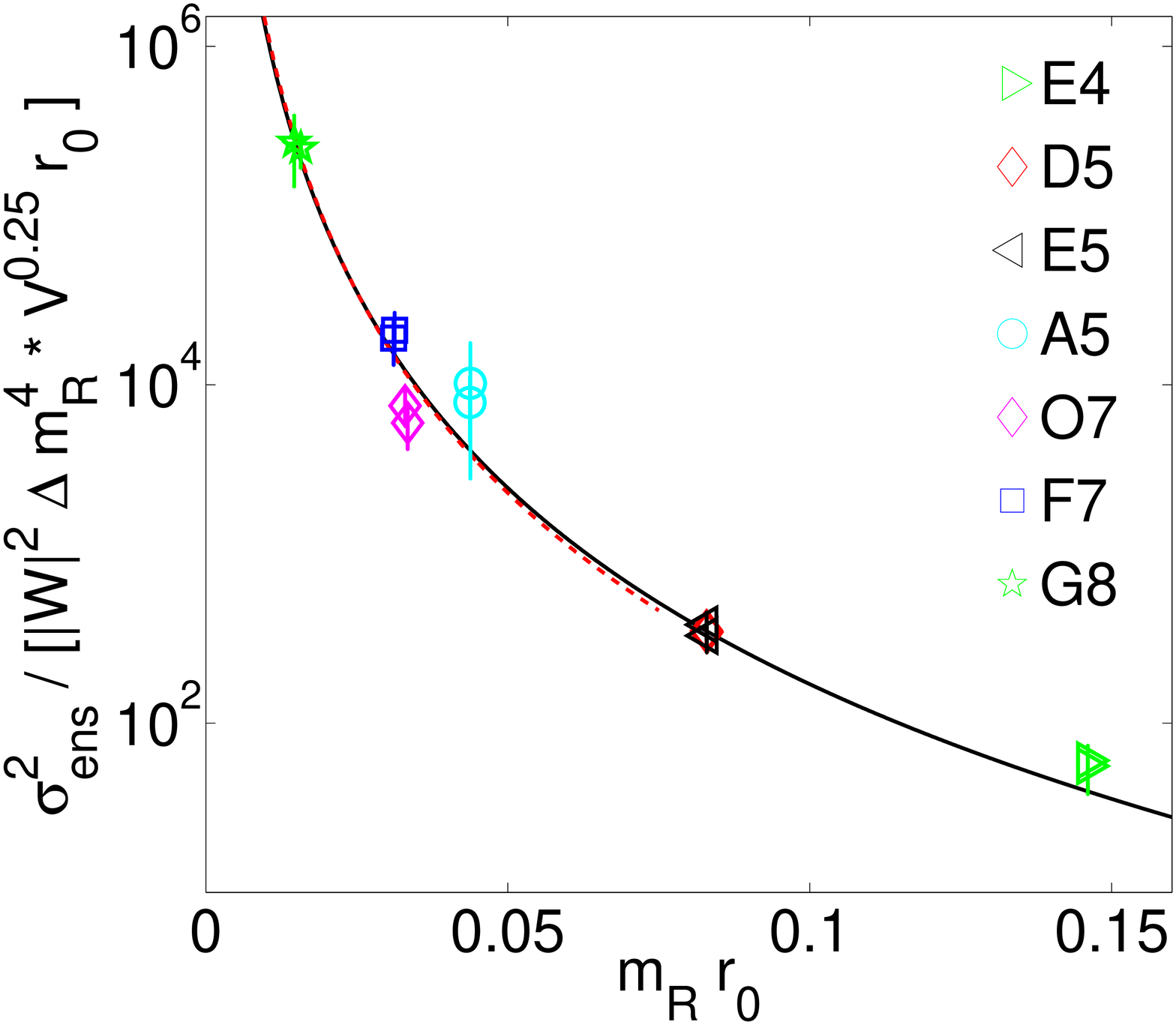} 
   \end{center}
   \end{minipage}
  \begin{minipage}[c]{.46\textwidth}
  \begin{center}
   \includegraphics*[angle=0,width=\textwidth]{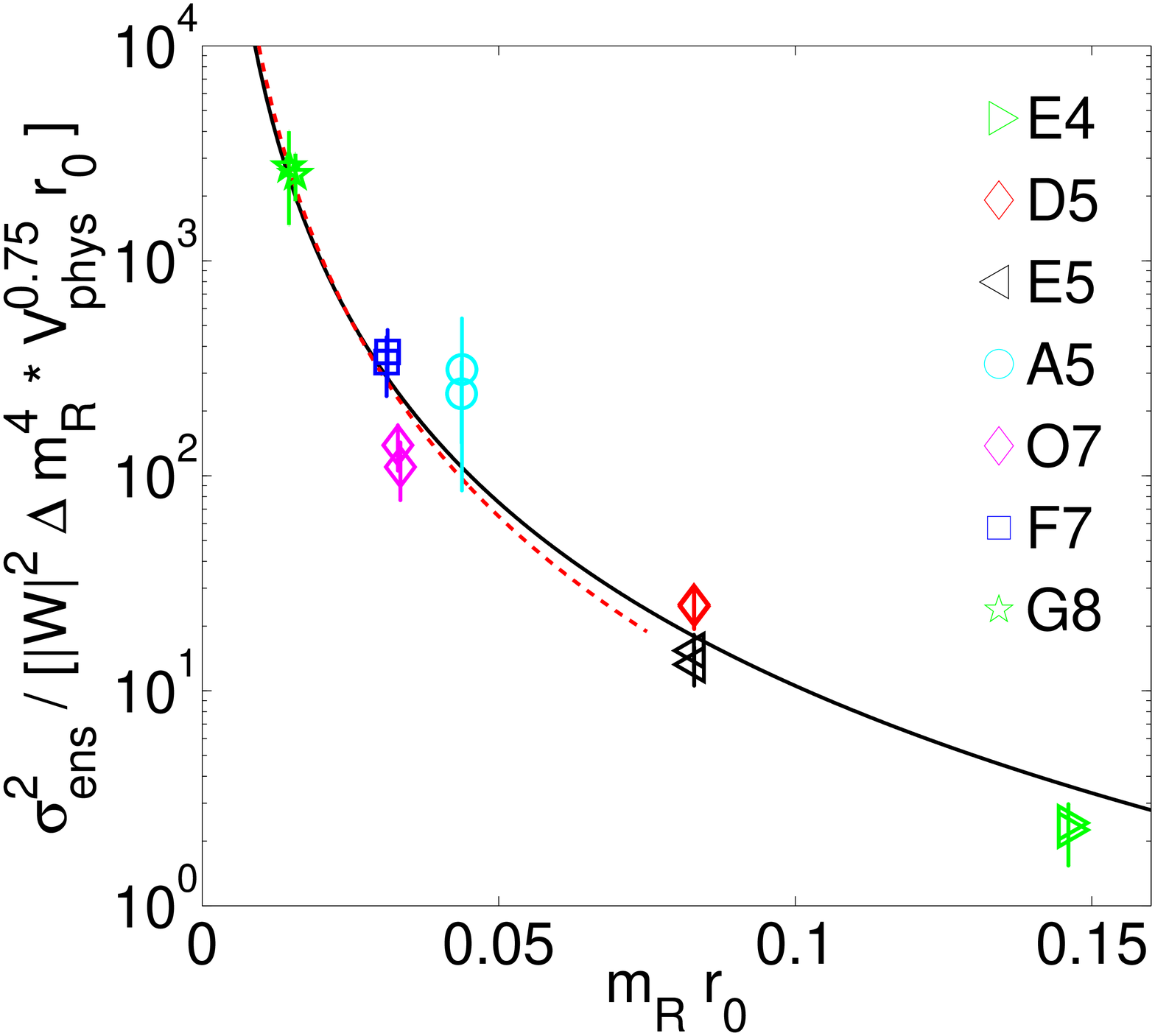} 
 \end{center}
 \end{minipage}
 \caption[Scaling of ensemble fluctuations]{\small The scaling of the ensemble fluctuations
 for different volume behaviors is shown. The left figure shows the quark mass behavior for a volume behavior of $\sqrt[4]{V}$
 and the right figure for a volume behavior of $V^{3/4}$.
}
\label{fig:eflc}
\end{figure}

By using the analyzed ensembles, listed in table \ref{tab:ens}, it
is possible to deduce numerically the scaling behavior of the fluctuations
in the quark mass and the volume.
However in the case of the stochastic fluctuations
the trace of the Wilson Dirac operator is known in chiral perturbation theory, e.g.~as in \cite{Giusti:2008vb}, by
$\langle \textrm{Tr} \frac{1}{\left(DD^\dagger\right)^2} \rangle \propto \frac{\Sigma V}{m_R^3}$.
The numerical analysis is consistent with this behavior.
It follows for the stochastic fluctuations (see fig.~\ref{fig:sflc})
\begin{equation}
 \sigma_{st}^2 \approx k_{st} \frac{\Delta m_R^4 V }{N_{inv} m_R^{r'}} \frac{1}{r_0^{r'}}
\end{equation}
by using the scale $r_0$ \cite{Sommer:1993ce} to form dimensionless quantities.
By fixing the volume behavior to $V$ we perform a fit (black, solid) to the quark mass behavior
by using the ensembles E4 (green, triangle), D5 (red, diamond), E5 (black, triangle),
A5 (cyan, circle), O7 (magenta, diamond), F7 (blue, square) and G8 (green, star). 
For each ensemble we compute $\sigma^2_{st}$ and $\sigma_{ens}^2$ 
for two values of $\Delta m = \Delta m_{max}/2,\Delta m_{max}$ (see tab.~\ref{tab:ens}).
The quark mass behavior is given by $r'=2.63(5)$.
For the red lines the quark mass behavior and 
the volume behavior is fixed to $V/m_q^3$ and only ensembles with pion masses $< 340 \textrm{MeV}$ are included.
The data show a good agreement with the expectation from
chiral perturbation theory for pion masses $< 340 \textrm{MeV}$.

The leading term of the ensemble fluctuations eq.~\eqref{eq:sens} 
is proportional to $\textrm{var}(\textrm{Tr} \, D^{-2})$.
Numerically we observe a weak volume dependence $V^q$ with $q<1$.
Similar to the stochastic fluctuations the ensemble fluctuations can be written as
\begin{equation}
 \sigma_{ens}^2 \approx k_{ens} \frac{\Delta m_R^4 V^q }{m_R^r} \frac{1}{r_0^{r-4-q}}.
\end{equation}
In general the simultaneous deduction of the volume and the quark mass behavior 
is difficult. A varied volume behavior changes simultaneously the mass behavior.
A good fit is given for a volume scaling of $q=0.25$ (see left figure \ref{fig:eflc}) which gives a mass behavior of $r=3.85(13)$
for all ensembles (black line) and $r=3.94(31)$ for ensembles with pion masses $< 340 \textrm{MeV}$ (red dashed line).
A weaker volume behavior is also supported by comparison of D5 and E5 ensembles, which gives $q \sim 0.46$.
However by assuming a similar quark mass behavior as in the case for the stochastic
fluctuations with $r=3$ the scaling in the volume is roughly given by $q \sim 0.75$. In the right figure \ref{fig:eflc}
we fixed the volume behavior to $q = 0.75$ which gives a mass behavior of $r=2.83(13)$ by including
every ensemble (black line) and $r=3.04(31)$ by including ensembles with pion masses smaller than 
$< 340 \textrm{MeV}$ (red, dashed line). We conclude that the volume behavior is given by $q \simeq 0.25 \ldots 0.75$ 
by a simultaneous variation of the quark mass behavior from $r \simeq 4 \ldots 3$. 

The cost of isospin reweighting can be estimated from the ratio
\begin{equation}
    \sigma^2_{st}(N_{inv})/\sigma^2_{ens} \sim \frac{k'_{st}}{k'_{ens}} \frac{(L m_{\pi})^2 L}{N_{inv} \cdot r_0} \qquad \quad \textrm{with} \quad\frac{k'_{st}}{k'_{ens}} = 1e-3 
\end{equation}
for $q=0.25$ and $r=4$. For the G8 ensemble follows  $N_{inv} \approx 200$ for a ratio of $0.1$. 

\section{Quark Masses}

\begin{figure}
\begin{center}
\includegraphics*[angle=0,width=\textwidth]{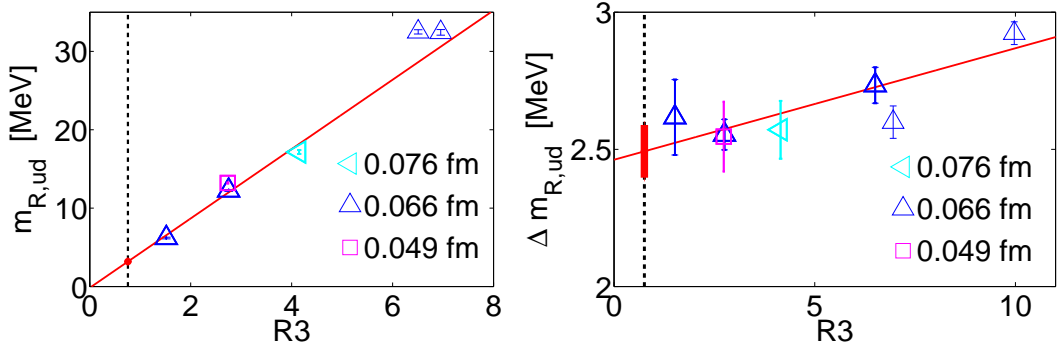}
\caption{\small The figures show the average quark mass $m_{ud}$ (left) and the mass splitting $m_d-m_u$ plotted versus $R_3$
after fixing $R_1$ and $R_2$. The masses are renormalized in the $\overline{\textrm{MS}}$--scheme at $2\,\textrm{GeV}$.}
\label{fig:qm}
\end{center}
\end{figure}

The continuum limit can be performed on a line of constant physics.
This line can be defined by keeping dimensionless ratios of 
physical quantities constant. These fix the bare mass parameters,
here, a quenched strange quark with $m_s$, the isospin mass splitting $\Delta m_{ud}$
and the average light quark mass $m_{ud}$. We take the ratios
\begin{equation}
R_1  = \frac{0.5(m^2_{K^0} + m^2_{K^\pm})}{(0.5(f_{K^0} + f_{K^\pm}))^2}   \quad,\quad R_2 =   \frac{m^2_{K^0} - m^2_{K^\pm}}{(0.5(f_{K^0} + f_{K^\pm}))^2} 
\quad \textrm{and} \quad R_3 = \frac{m_{\pi^\pm}^2}{(0.5(f_{K^0} + f_{K^\pm}))^2}
\end{equation}
with the meson masses, the pion $m_{\pi^\pm}$, the neutral kaon $m_{K^0}$ and the charged kaon $m_{K^\pm}$ 
and the kaon decay constants $f_{K^0}$ and $f_{K^\pm}$.
The physical values of the ratios are taken from \cite{Aoki:2013ldr} and we assume $0.5(f_{K^0}+f_{K^\pm})=155 \;\textrm{MeV}$.
Now, the strategy is to use $R_1$ to fix $m_s$, which is done in \cite{Fritzsch:2012wq}
and $R_2$ to fix the isospin splitting $\Delta m_{ud}$.
Afterwards the ratio $R_3$ is used to extrapolate
the light quarks towards the physical limit.

We measure the PCAC mass on the analyzed ensembles (see tab.~\ref{tab:ens}) 
and convert them into the $\overline{\textrm{MS}}$-renormalization scheme.
The dimensionless ratios $R_2$ and $R_3$ are given in the lowest order chiral perturbation 
theory up to $\mathcal{O}(\Delta m_{ud}^2,m_{ud}^2)$ by $R_2  = \frac{B}{F^2} \, \Delta m_{ud} (1 + C m_{ud})$ and  $R_3 = \frac{2 B}{F^2} \, m_{ud}$
with constants $B$, $C$ and $F^2$.
Now, it is possible to perform extrapolations towards the physical point in the light quark masses.
For the average light quark mass this is shown in the left figure of fig.~\ref{fig:qm} by assuming $ m_{ud}(R_3) \approx a_1 R_3 $.
By using the F7 and G8 ensemble the average light quark mass at the physical point at finite lattice spacing 
of $a=0.066 \; \textrm{fm}$ is given by $m_{ud,R} = 3.19(11) \; \textrm{MeV}$.
For the mass splitting in the light quarks (see right plot in figure \ref{fig:qm}) we assumed $ \Delta m_{ud}(R_3) \approx b_0 + b_1 R_3$.
By using the data of the E5, F7 and G8 ensemble it follows for
the mass splitting $\Delta m_{ud} = 2.49(10) \; \textrm{MeV}$ at finite lattice spacing $a=0.066 \; \textrm{fm}$.

The isospin effects enter the observable by the isospin reweighting factor
which scales proportional to $\Delta m_{ud}^2$. In the case of the PCAC mass
the statistical error is too big compared to the effect of the isospin reweighting correction.
A determination of this effect is only possible for larger statistics.
Neclecting the sea quark effects proportional to $\Delta m_{ud}^2$,
by setting the isospin reweighting factor to one,
the isospin quark mass splitting is given by $\Delta m_{ud} = 2.52(10) \; \textrm{MeV}$.
However the isospin reweighting effects increases for smaller quark masses 
and we want to reduce the statistical error to figure out the isospin 
effects for example in the pion mass.  
In order to perform a continuum limit the statistics has to be increased
and other ensembles have to be included. 

\section{Conclusion}

Isospin mass reweighting needs a moderate numerical effort. The analysis shows 
that the cost scales with $(L M_{PS})^2 L$ for a volume scaling
of the ensemble fluctuations with $\sqrt[4]{V}$ and is
around $200$ inversions of the Dirac operator for the G8 ensemble 
which has a pion mass of $192$ MeV at a volume of $V/a^4= 128 \times 64^3$.
By using the introduced dimensionless ratios $R_1$, $R_2$ and $R_3$
it is possible to extract the light quark masses.
The isospin mass spliting is $\Delta m_{ud} = 2.49(10) \; \textrm{MeV}$
and the average quark mass is $m_{ud,R} = 3.19(11) \; \textrm{MeV}$
at finite lattice spacing of $a=0.066 \; \textrm{fm}$.
Although a more careful analysis is needed to extract
competitive numbers it shows that the tuning
conditions are suitable to extract the light quark masses.
In order to extract continuum physics
the statistics has to be improved and QED-effects have to be included.
A software package for mass reweighting \cite{soft:OpenQCDwithM} (see also \cite{Leder:2014})
is publicly available in the framework of the \textit{openQCD} code\cite{soft:OpenQCD}.

\end{document}